# Domain specific ontologies from Linked Open Data (LOD)

Rosario Uceda-Sosa, Nandana Mihindukulasooriya
IBM TJ Watson Research Center
Yorktown Heights, NY, USA
rosariou@us.ibm.com

Atul Kumar, Sahil Bansal, Seema Nagar
IBM Research - India
Bangalore, India
Kumar.atul@in.ibm.com

## ABSTRACT

Logical and probabilistic reasoning tasks that require a deeper knowledge of semantics are increasingly relying on general purpose ontologies such as *Wikidata* and *DBpedia*. However, tasks such as entity disambiguation and linking may benefit from domain specific knowledge graphs, which make it more efficient to consume the knowledge and easier to extend with proprietary content. We discuss our experience bootstrapping one such ontology for IT with a domain-agnostic pipeline, and extending it using domain-specific glossaries.

## 1 INTRODUCTION

Ontologies are being relied upon to support logical and probabilistic reasoning in tasks like information extraction and query answering, especially now that Deep Learning (DL) research moves towards Neuro-Symbolic approaches [15] which require a deeper understanding of semantics. However, except for few domains such as life sciences, finding an open source, domain-specific ontology with a rich conceptualization is often difficult and creating an ontology from scratch, very expensive.

Our work was initially motivated by the need for an ontology about IT operations to help downstream applications. To the best of our knowledge, there are no public IT domain ontologies available for reuse, thus the authors decided to create one, leveraging open data while extending it with client-specific information (e.g., public user forums, manufacturer-provided documents, Wikipedia, etc.)

Original use cases for this work include generating facets to facilitate the search of customer support documents [31] or improving the performance of terminology extraction and ranking in knowledge induction scenarios [12], [7], [37]. While unsupervised generation of knowledge graphs from text is promising, there are still performance issues that hinder their use [26], [6]. This may be due in part to noisy terminology and term overloading (e.g., homonyms and homographs), which cause the system to struggle when finding clusters of meaning from sparse specialized terms. On the other hand, Linked Open Data (LOD) resources, especially those heavily used and curated (e.g., Wikidata) are generally reliable, rich (91 million entities) and up-to-date.

Here, we discuss our experience with the construction and evaluation of an IT operations ontology which is publicly available for evaluation and further experimentation under an open license in https://github.com/IBM/ITOPS-ontology.

## 2 RELATED WORK

Several general knowledge graphs have been released in the public domain over the last few decades. These resources, ranging from WordNet, released in 1995, to DBPedia, YAGO and Wikidata are well known in the Semantic Web community. There are also domain specific datasets such as ResearchSpace [36], a cultural heritage knowledge graph, UMLS [34], a medical language ontology, GeneOntology [16], a gene ontology resource, SNOMED CT [38], a commercial clinical terminology database, and Yidu Research's medical knowledge graph [43].

The Linked Open Cloud dataset[25] contains 1,269 datasets with 16,201 links between them. Similarly, Linked Open Vocabularies [40] provides a comprehensive catalogue of open vocabularies. We have performed a thorough search of these resources but, to the best of our knowledge, there is no public ontology for IT Operations rich enough for the use cases mentioned above.

A pipeline for ontology construction, combining Semantic Web and ML/DL technologies involves an extensive body of knowledge in the areas of ontology design, semantic web, dynamic graph generation from text, entity relation and linking, graph embedding and propositionalizations and ontology evaluation areas, among others. For the sake of brevity, we focus on the areas that are related to our approach.

We reviewed the more traditional construction process of other specialized knowledge graphs [13], [39], [27], [8]. Unfortunately, these don't offer a clear road map to automatically build specialized ontologies for an arbitrary domain. We also looked into well-known efforts to induce an ontology from text, [22], [20], [35] but without fully reliable results as of yet.

There has been much work to leverage LOD graphs [32], generate Wikidata *IS-A* hierarchies [3] or to classify knowledge organization systems [41]. None of these approaches generates rich ontologies, though. Also, there has been work [29] to integrate existing curated ontologies with induced knowledge graphs. Unfortunately, their method works best in domains such as biochemistry or life sciences, where structured databases with terms unique to the domain already exist.

Reference ontologies have also been identified as a viable tool to simplify and speed up the construction of application ontologies. However, most reference ontologies do not capture as of yet the existing upper ontologies of popular LOD graphs [42], [10], [2].

## 3. DOMAIN-AGNOSTIC PIPELINE FOR DOMAIN ONTOLOGIES

Even though the scope of this paper is limited to ITOPS, we wanted the construction process itself to be domain-agnostic so it can be leveraged in other domains, like oil and gas or finance. For that purpose, we start with a minimal T-Box, instantiated by a General Library of Objects (GLO).

In contrast to [11, 19, 30, 33] GLO is not a formal upper ontology and foregoes abstract concepts (e.g., *SelfConnectedObject*, *PureSubstance*) in favor of those that describe human activity and events, already familiar to domain experts. We construct it from Wikidata top-level concepts [1] cross referenced to documents describing stakeholders, locations, resources and activities resulting on 74 concepts that cover 37 million Wikidata entities.

We also define a set of high-level relations among them, like *hasPart* (wdt:P527) or *equivalentClass* (wdt:P1709). The backbone of the GLO taxonomy is *subConceptOf*, which subsumes *wdt:P279* (subclass of) and *wdt:P31* (instance of). This ISA overloading [18] does not affect the quality of most user queries. Besides, enforcing a strict T-Box creates practical problems in dynamic ontologies as entities like a computer model, a medication or an illness, can be both viewed as instances or classes, depending on the application.

From the initial GLO entities, our ITOPS pipeline consists of three automatic steps described below. ITOPS (as well as other ontologies produced by our pipeline) belongs to the OWL RL dialect of OWL-2, whose queries and optimization tasks can be efficiently computed[2].

*ITOPS S1, Extracting Explicit LOD.* We start with a set of positive seed concepts, as well as an optional set of negative concepts that will not be populated. These seed concepts can be defined by users or extracted from the text.

The next step is to automatically extract a Wikidata subgraph based on those seed concepts and map it [10] into the GLO entities. From the positive seed concept list $Con = \{C_1, ..., C_n\}$ and the negative concept list $NCon = \{NC_1, ..., NC_n\}$ a graph $GC = \{V, E\}$ is extracted. Here $E$ is the set of sub-properties of *Wikidata Property* (wd:Q18616576) and $V$ is the set of entities (Wikidata items) $V_i$ so that $V_i$ (*wdt:P279* OR *wdt:P31*) $* C_j$ and the opposite holds wrt $NCon$ vertices. These generated vertices are instances related through *glo:subConceptOf*. Other relations among members of $V$ are also added. If a relation's object is another entity not in $V$, data is brought in as a text field. ITOPS S1 has 47,300 entities, of which 2,000 of them have subconcepts of their own, plus 212 domain-specific relations.

*ITOPS S2, Extracting Implicit LOD.* Despite being one of the largest LOD datasets with ~91 million entities, Wikidata is far from being complete.

We use the category tags in Wikipedia articles to expand S1. Such tags contain information that is not explicitly in Wikidata so, for each S1 entity we get its associated Wikipedia category. Not all categories are relevant to IT, though. Take *computer keyboard* (Q250); its Wikipedia page is tagged with four categories *computer peripherals*, *computing input devices*, *flexible electronics*, and *game control methods*. While the first two are IT-specific categories, the latter two are less relevant.

To find useful categories, we look up their members and their Wikidata entries along with their parent classes. Then we measure the heterogeneity of member entities and their overlap with S1. For example, entities in *computer peripherals* are instances of classes that subclass *peripheral equipment* (Q178648), while entities in *flexible electronics* have parents from many different classes, all far from each other in the ISA hierarchy. Furthermore, a large portion (75%) of entities in 'computer peripherals' overlap with the initial partial forest while for 'flexible electronics' it is only 9%.

The metrics we use to identify useful categories include a normalized number of classes (the number of classes divided by the number of entities), average path length (in the ISA hierarchy) between each class pair average depth from the common ancestor, class overlap and entity overlap with ITOPS S1. With these metrics as features and a manually curated sample of positive (domain-specific) and negative (generic) categories, we train a binary classifier to decide categories relevant to IT and with the right granularity. This step, S2, identified 5,589 new entities. After eliminating low scored entities, 4,752 entities were added to S1.

*ITOPS S3, Domain Specific Data Extensions.* We extend S2 with public glossaries in the IT domain, as proprietary client data could not be used in this paper. S3 adds glossary terms by finding the right parent in S2 by *subConceptOf*. There is prior work to add terms to an ontology [46], [1], [44] but these methods either need a reference term or a domain-specific corpus so they cannot be applied in our scenario.

Our approach propositionalizes S2 using a predicate-based approach as in [5] and then encode each propositionalization into a 500-dimensional embedding vector. The choice of a predicate-based approach is driven by having a definition-like representation of an entity so that similarity computation with the glossary terms is possible. It also enables us to have useful definitions for the ontology entities. Similarly, we also encode each glossary term to an embedding vector from the glossary definition. We have used USE[3] for encoding the entities and the glossary terms. For each glossary term, we find the $topK$ most similar entities in S2 using the encoded vectors. We experimented with $topK$ as 5, 10 and 15 and found that 5 and 10 give similar and best results. Using the $top$5 most similar entities, we find the parent of the glossary term to whom the glossary term is going to be attached as *glo:subConceptOf*.

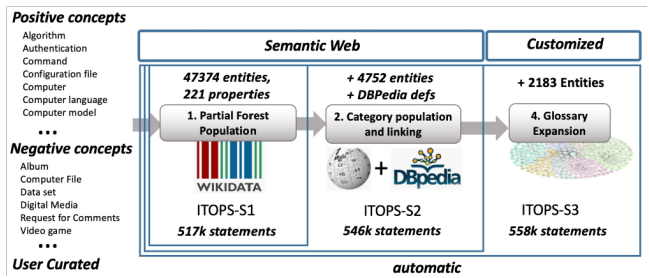

Figure 1: Automated ontology construction pipeline from seed concepts

---

[1] https://www.wikidata.org/wiki/Wikidata:WikiProject_Ontology/Toplevel_ontology_list

[2] https://www.w3.org/TR/owl2-profiles/ref-owl-2-specification

*Evaluation of ITOPS S3.* We conducted experiments to evaluate the efficacy of our approach by performing both a manual and an automated validation. We use four glossaries (1) DB2 [23], (2) Cloud Object Storage [24], (3) Lenovo [28] and (4) IT Terms [9], coming from diverse sources for extending S2 for the manual evaluation. After removing duplicate terms, the total number of terms in all the glossaries are 2, 181. For the automatic evaluation, we select a certain percentage of the leaves to delete from S2 and add them back using our approach. We experiment with 5%, 10% and 15% of the leaves.

For the manual evaluation, we add glossary terms to S2 using our approach and asked three Subject Matter Experts (SMEs) to manually evaluate the parent of the glossary terms, assigning a label 1 if the parent was correct, 0 if the parent was wrong, 2 if the parent was reasonably related and −1 if the parent was the root node of S2, meaning there's no meaningful parent information.

We achieve an inter-annotator agreement of 0.71 with Fleiss' Kappa in the manual evaluation. Table 1 shows the details of the annotator agreement. We observe that some of the terms are labelled as 0 because S2 does not have the entity (parent) to which they should be attached. E.g., *rollout, hole, update hole* from the DB2 glossary are attached to glo:*sql_keyword*. Instead, they should have been attached to *database_keyword*, which is not present in S2 (140 such instances in total). Also, a considerable number of terms are labelled as −1 (parent is the root node in S2) because their parent concepts are not present in S2. Around 20% from a total of 38.51% terms labelled with −1 are actually placed correctly so our approach is able to place around 50% of the terms at their correct position, and 20% are placed at a related position.

Table 1: Distribution of Labels

| Annotator | % Label 1 | % Label 0 | % Label 2 | % Label -1 |
|---|---|---|---|---|
| 1 | 25.86 | 15.59 | 21.04 | 38.51 |
| 2 | 16.78 | 25.22 | 20.45 | 38.51 |
| 3 | 23.15 | 18.15 | 20.49 | 38.51 |

In the automated evaluation we remove certain nodes from the graph and try to add them using our algorithm. This means we have ground truth labels and can compute the distance (path length by the *subConceptOf* property) between the assigned parent and the ground truth parent as a quantitative performance metric. When the assigned parent is in a different *subConceptOf* hierarchy then there is no reachable path. The results for different percentages of leaves deletion are shown in Table 2 where the distance is computed as defined earlier, and *NotReachable* corresponds to the entities for which the ground truth parent is not reachable from the identified parent. Our approach performs well with more than 60% of the leaves attached within 2 hops away. A possible reason for around 30% *NotReachable* entities could be the fact that in the current form of S2 if we consider the sub-trees under the root node, *glo:IT_Entity*, there are several hundred subtrees with the tree height being <= 2. For each such sub-tree, once we remove the leaf nodes, the remaining nodes might not have enough representative information of the subtree.

Table 2: Automatic Validation Statistics

| Distance | 5% leaves | 10% leaves | 15% leaves |
|---|---|---|---|
| 0 | 37.5 | 36.7 | 36.4 |
| 1 | 13.1 | 15.1 | 12.6 |
| 2 | 10.6 | 8.3 | 9.0 |
| 3, 4, 5 | 11.9 | 9.9 | 11.9 |
| *NotReachable* | 26.9 | 30 | 30.1 |

*The ITOPS Resource.* ITOPS is published following the FAIR principles for findability, accessibility, interoperability, and reusability. The resource is registered in Zenodo[4] and GitHub[5]. The canonical persistent URI for the ITOPS ontology resource is *http://doi.org/10.5281/zenodo.3942013*. It can be downloaded under businessfriendly Apache 2.0 open license. The resource is compliant to the Semantic Web standards such as OWL, RDF, and Turtle. A DCAT dataset description is available with data.

Regarding sustainability and maintenance, the original authors are currently maintaining and improving it based on the feedback from downstream application users. For reusability, documentation and a tutorial are available in the GitHub repo providing details on how it was constructed, a schema diagram and how to use and query it.

## 3 THE EVALUATION OF ITOPS

There are several dimensions that help in measuring and evaluating an ontology [21], [17], [14]. In our case, ITOPS is both derived from and extends a larger ontology, Wikidata, so we assess whether it is more complete (coverage) or less noisy (conciseness) than the general Wikidata for the IT domain. We also want to know how relation-rich it is (complexity) and whether it is useful to terminology ranking.

We selected the titles of 100,000 documents of the TechQA Dataset [4], each of which describes a thread corresponding to a user request for help on subjects ranging from hardware to networking questions.

---

[3] https://tfhub.dev/google/universal-sentence-encoder/4
[4] https://zenodo.org/record/3942013
[5] https://github.com/IBM/ITOPS-ontology/releases/tag/v0.1

|  | WD | S1 (Forest) | WD+ S1 | S2 (Categories+) | S3 (Glossaries) |
|---|---|---|---|---|---|
| Coverage (percentage) | 15.6 (first 3) 29.2 (first 5) | 44.4 | 51.7 (first 3) 62.3 (first 5) | 46.9 | 47.2 |
| Conciseness | 6.30 | 1.39 | N/A | 1.37 | 1.35 |

| Complexity | S1 (Forest) | S2 (Categories+) | S3 (Glossaries) |
|---|---|---|---|
| Entities/Relations | 47374/221 | 52,126/221 | 54309/221 |
| Statements (all/minus inheritance) | 546,572/386,223 | 475,932/397,530 | 558,028/403,259 |
| Connectivity (all/minus inheritance) | 11.53/8.1 | 9.13/7.62 | 10.27/7.42 |

Figure 2: Evaluation of ITOPS Stages

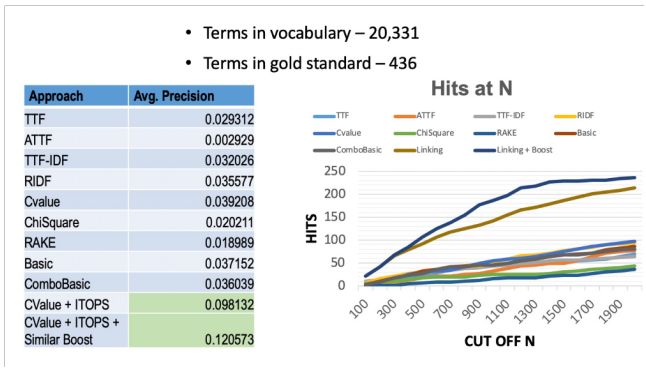

Figure 3: Terminology Ranking Results

Next, we look at the list of terms returned by the Wikidata Term Search Service [6] and examine the position of the first IT-related term. For example, the query for *baseline* returns a list where the first IT term (software release, alias *baseline*, Q20631656) is in the 4th place.

Many of the terms (85%) in the titles were too generic (e.g., *number*, *delay*). However, out of the IT terms, Wikidata returned only 15.6% in the first three and 29.2% in the first five positions. ITOPS-S1, has a coverage of 44.4%, but by using ITOPS to give precedence to terms with the shortest paths to the 2000 non-leaf entities, we are able to place 51.7% in the first three and 62.3% in the first five positions (WD+S1 Column), showing that ITOPS prioritizes the classification of domain-specific terms.

As for determining conciseness, we look at the average length of the list of terms returned where there is a correct (related to the IT domain) term, only counting non-empty lists, so as not to mix conciseness with coverage. As we extend it with S2 and S3, ITOPS becomes more fitted to the application (see Fig. 2).

The second table in Fig. 2 measures other aspects of the ITOPS graphs, like the size of *productive* statements by excluding metadata (provenance, labels, etc.) and the average entity connectivity in the graphs' deductive closure (third row). As expected, the general connectivity goes down in S2 and S3, since we're not adding new properties.

Lastly, we present a use case where the ITOPS was used to boost terminology ranking performance. The corpus used was a set of 4,000 technical troubleshooting documents in the personal computer domain, which are relatively short and contain symptoms and solutions for a specific customer problem. For example, one such document may provide the solution for "Webcam not working" in a given laptop model. Using an internal service [12], we automatically extracted a vocabulary of 20,331 terms from these documents. Ranking these documents by keywords helps users search for the most relevant solutions.

To evaluate the ranking quality, a manual gold standard of 480 terms was created. Average precision was calculated for the ranked list of each method as the evaluation metric, where the terms in the gold standard were considered relevant and the rest are negative.

Existing approaches for terminology extraction and ranking didn't perform well on this corpus because the documents are short and the technical text contains mostly lists, logs, or commands. Also the important terms, like computer models, have low frequency. The table on the left of fig 3 shows the average precision of the ranked lists returned by each method (details on the terminology ranking approaches in [45]). The last row (*CValue + ITOPS*) shows the improvement by leveraging ITOPS. Similarly, the chart on the right shows hits (relevant terms) at each cutoff N in the ranked list returned by each method, clearly demonstrating the value of ITOPS in terminology ranking.

## 4 CONCLUSIONS

We have presented an ontology for IT Operations that will be useful downstream applications such as dynamic faceted search [31] or knowledge induction from text [12], [7], [37]. Our motivation was to leverage good quality, actively curated and widely-used knowledge graphs such as Wikidata to automatically absorb the frequent changes that knowledge in technical domains often goes through. At the same time, we demonstrated that we could add specific knowledge sources such as domain glossaries reliably.

---
[6] https://www.wikidata.org/w/api.php